\documentclass[aps,prl,twocolumn,superscriptaddress,nofootinbib]{revtex4-2}

\usepackage{amsmath,amssymb}
\usepackage{graphicx}
\usepackage{bm}

\newcommand{\Sthree}{S^3}

\begin{document}

\title{Coarse graining from within: Wilson--Fisher universality on $S^3$}

\author{Alfio M. Bonanno}
\affiliation{INAF – Osservatorio Astrofisico di Catania, Via S. Sofia 78, 95123 Catania, Italy}
\affiliation{INFN, Sezione di Catania, Via S. Sofia 64, 95123 Catania, Italy}

\author{Renata Ferrero}
\affiliation{Institute for Quantum Gravity, 
Friedrich-Alexander-Universität Erlangen-Nürnberg, Staudtstr. 7, 91058 Erlangen, Germany}

\author{Giovanni Oglialoro}
\affiliation{INAF – Osservatorio Astrofisico di Catania, Via S. Sofia 78, 95123 Catania, Italy}
\affiliation{INFN, Sezione di Catania, Via S. Sofia 64, 95123 Catania, Italy}
\affiliation{Dipartimento di Fisica e Astronomia, Universit\`a di Catania, Via S. Sofia 64, 95123, Catania, Italy}

\date{\today}

\begin{abstract}
Wilsonian renormalization is usually formulated in momentum space, but on curved backgrounds momentum shells have no invariant meaning.
We replace them by an intrinsic spectral cutoff, ordering modes by the covariant Laplacian and setting the cutoff resolution by the system size in renormalization group (RG) units.
For a scalar field on $S^3$, this yields a covariant, momentum-free RG flow whose trace is an exact sum over spherical harmonics.
The standard flat-space flow is recovered when the sphere is large compared with the coarse-graining scale.
As a nontrivial test, the compact spectral flow realizes Wilson--Fisher universality without momentum shells: the interacting fixed point survives at finite resolution, has one relevant direction, and approaches its flat-space counterpart smoothly, with critical exponents only weakly affected by the compact spectrum.
\end{abstract}

\maketitle

\paragraph*{Introduction.} Wilsonian renormalization provides a framework for describing scale dependence in physical theories.
  It accounts for the emergence of universality classes and for the insensitivity of long-distance physics to microscopic details~\cite{Kadanoff:1966wm,Wilson:1971dc,Wilson:1971bg,Wilson:1972zz,Wilson:1974mb,ZinnJustin:2002ru,Pelissetto:2000ek}.  In its most familiar form this idea is expressed as a momentum-shell construction: one lowers a cutoff scale $k$ and integrates fluctuations according to their momenta.

Exact renormalization-group equations give this picture a functional, nonperturbative realization~\cite{Wegner:1973,Polchinski:1983gv,Wetterich:1992yh,Morris:1993qb,Berges:2000ew,Delamotte:2012nb,Dupuis:2020fhh}.  In the functional RG formulation \cite{Wetterich:1992yh, Morris:1993qb}, Wilson--Polchinski flows \cite{Polchinski:1983gv}, and proper-time flows \cite{Liao:1996pd} differ in formulation, but share the interpretation of $k$ as a resolution scale.  In exact treatments universal quantities are regulator independent, while in truncations such as the local-potential approximation (LPA) a residual scheme dependence remains and provides a measure of the approximation~\cite{Hasenfratz:1986dm,Tetradis:1993ts,Morris:1994ie,Litim:2001up,Litim:2002cf,Litim:2010tt,Canet:2003qd}.  Proper-time flows, which descend from the Schwinger representation of the one-loop effective action~\cite{Schwinger:1951nm}, provide a particularly simple setting for studying this issue and have been successfully used in critical phenomena~\cite{Liao:1996pd,Bonanno:2000yp,Bonanno:2004sy}.

The momentum-shell picture relies implicitly on translation invariance.  In flat Euclidean space, plane waves diagonalize the kinetic operator and momenta consequently provide invariant labels of fluctuations.  Once this symmetry is absent, however, a momentum shell is no longer a coordinate-independent object.  This obstruction is generic in settings without global translation invariance, from finite-size critical phenomena~\cite{Fisher:1972zza,Privman:1984zz} and quantum fields on curved or cosmological backgrounds~\cite{BirrellDavies:1982,Hollands:2002ux,ParkerToms:2009} to quantum gravity, where no fixed background translation group should be assumed~\cite{Reuter:1996cp,Kiefer:2007ria,Thiemann:2007pyv,Reuter:2019byg}.

The Wilsonian notion of scale must then be reformulated covariantly, with momentum shells replaced by an intrinsic ordering of fluctuations.
Finite-volume critical phenomena already indicate what this replacement should capture: the low-lying spectrum and the zero mode are physical degrees of freedom, while the approach to the thermodynamic limit is itself part of the universal scaling behavior~\cite{Fisher:1972zza,Brezin:1981gm,Brezin:1985xx,Privman:1984zz,Cardy:1988, Chen_2000}.  Accordingly, an RG flow on a compact space should keep the discrete spectrum explicit. The role of low modes and finite size is well known from finite-size scaling and finite-volume field theory, and has also been studied in  functional RG settings  treating it as a finite-size correction to a continuum momentum integral \cite{Fister:2015eca} or on spherical and hyperbolic spaces \cite{Benedetti:2014gja}.
In the present construction, the cutoff must  be defined relative to the geometry that supplies the spectrum, not only to analyze infrared curvature effects, but also to define the Wilsonian coarse graining intrinsically.

We  advance a spectral coarse graining as a covariant replacement for the momentum-shell cutoff. The intrinsic fluctuations of a field on a Riemannian manifold are the eigenfunctions of the  covariant kinetic operator.  Their eigenvalues provide a coordinate-independent measure of inverse wavelength, while their degeneracies count the number of independent fluctuations at each spectral level.  This is the same principle that underlies heat-kernel methods, spectral sums, and covariant functional traces, where invariant quantities are systematically built from functions of Laplace-type operators rather than from coordinate momenta~\cite{DeWitt:1965,Buchbinder:1992rb,Vassilevich:2003xt,Gilkey:1995,Kirsten:2001,Chavel:1984}.  Here, however, the spectrum is not only a calculational device: it defines the Wilsonian ordering itself.  Related spectral regularizations have appeared in the form of $N$-cutoffs, where the regulator directly restricts the mode content rather than coordinate momenta~\cite{Becker:2020mjl,Becker:2021pwo,Banerjee:2023ztr,Ferrero:2024yvw,Ferrero:2025ugd}.

In this Letter we develop this framework in the simplest setting where the relevant spectral data are known explicitly: a single real scalar field on a  three-sphere $\Sthree$~\cite{Camporesi:1994ga, Benedetti:2014gja}.
We construct a proper-time  RG flow and and use the Wilson--Fisher fixed point as a diagnostic of whether the compact spectral flow preserves Wilsonian universality without momentum shells.

\paragraph*{An intrinsic cutoff at work.}
On $\Sthree$, the role of momentum is taken over by the spectrum of the covariant Laplacian.
 The space is compact, homogeneous, and curved.  It has no global translation group, and hence no coordinate-independent momentum shell, but its scalar spectrum is known exactly~\cite{Camporesi1990,Camporesi:1994ga}.  For a sphere of radius $a$,
\begin{equation}
-\nabla^2Y_{nlm}
=
\frac{n^2-1}{a^2}Y_{nlm},
\qquad
D_n=n^2,
\qquad
n=1,2,\ldots .
\label{eq:spectrum}
\end{equation}
The eigenvalues order fluctuations invariantly, while the degeneracies $D_n$ count the number of modes at each spectral level.

What remains is to relate this spectral ordering to the physical resolution of the cutoff. The integer $n$ is dimensionless; it labels a position in the spectrum, but not a physical coarse-graining scale.  The cutoff must compare the wavelength of a mode with the RG length $\ell_k\sim k^{-1}$.  This comparison introduces the dimensionless variable
\begin{equation}
\tau=ka .
\label{eq:tau}
\end{equation}
It counts the angular harmonics resolved by the RG flow, since
\begin{equation}
\frac{n}{a}\lesssim k
\qquad\Longleftrightarrow\qquad
n\lesssim \tau .
\end{equation}
 We call this an intrinsic cutoff: its physical meaning is fixed by comparison with the geometry on which the field propagates. Its physical calibration comes from within the system.

In a compact space, this comparison is encoded in the dimensionless resolution $\tau$. Large $\tau$ corresponds to a local regime in which many harmonics are sampled and the spectrum is effectively dense.  Values $\tau=O(1)$ correspond to a pure compact regime in which the flow resolves only the first few global modes, including the homogeneous mode.  Finite-size and curvature effects therefore enter through the spectral trace itself, not as corrections added to a flat-space momentum integral.

\paragraph*{Wilson--Fisher universality at finite resolution.}
The Wilson--Fisher fixed point provides a  test of the construction.  It is the canonical non-Gaussian fixed point of three-dimensional scalar field theory and controls the Ising universality class~\cite{Wilson:1972zz,Pelissetto:2000ek}.  If the intrinsic spectral cutoff implements Wilsonian coarse graining correctly on $\Sthree$, the fixed point should persist at finite $\tau$ and connect smoothly to the usual flat-space fixed point when the spectrum becomes dense.  This gives a direct test of Wilsonian universality without momentum shells.

We work in the LPA for the scale-dependent effective action $\Gamma_k$, keeping the kinetic term fixed and encoding the RG flow in the potential $U_k$:
\begin{equation}
\Gamma_k[\phi]
=
\int_{\Sthree} d^3x\sqrt{g}
\left[
\frac{1}{2}g^{\mu\nu}\partial_\mu\phi\partial_\nu\phi
+
U_k(\phi)
\right].
\label{eq:lpa}
\end{equation}
For a constant background field, the Hessian entering the flow is
\begin{equation}
\Gamma_k^{(2)}[\phi]=-\nabla^2+U_k''(\phi).
\label{eq:hessian}
\end{equation}
We evolve $\Gamma_k$ with the proper-time functional RG flow~\cite{Liao:1996pd,Bonanno:2000yp,Bonanno:2004sy,Litim:2010tt,Bonanno:2019ukb}
\begin{equation}
\partial_t\Gamma_k[\phi]
=
-\frac{1}{2}
\int_0^\infty\frac{ds}{s}
\,\rho_k^{(m)}(s)\,
{\rm Tr}\,
\exp[-s\Gamma_k^{(2)}[\phi]],
\label{eq:ptflow}
\end{equation}
using $t=\log k$, and the regulator family
\begin{equation}
\rho_k^{(m)}(s)
=
-\frac{2}{\Gamma(m)}
(k^2s)^m e^{-k^2s}.
\label{eq:regulator}
\end{equation}
The parameter $m$ controls the localization of the proper-time threshold. In three dimensions, the spectral sums appearing below are finite for $m>3/2$.

Inserting the spectrum \eqref{eq:spectrum} into the trace gives, for any function $F$,
\begin{equation}
{\rm Tr}\,F(-\nabla^2)
=
\sum_{n=1}^{\infty}
n^2
F\left(\frac{n^2-1}{a^2}\right).
\label{eq:trace}
\end{equation}
Dividing by the volume $V_{\Sthree}=2\pi^2a^3$ and performing the proper-time integral, one obtains the exact compact spectral flow
\begin{equation}
\partial_tU_k(\phi)
=
\frac{(ka)^{2m}}{2\pi^2a^3}
\sum_{n=1}^{\infty}
\frac{n^2}
{\left[n^2+\widetilde M^2(\phi)\right]^m},
\label{eq:dimensionfulflow}
\end{equation}
where
\begin{equation}
\widetilde M^2(\phi)
=
a^2k^2
\left(
1+\frac{U_k''(\phi)}{k^2}
\right)-1 .
\label{eq:Mtilde}
\end{equation}
The shift by $-1$ is the curvature contribution in the scalar spectrum on $\Sthree$.  Equation~\eqref{eq:dimensionfulflow} is the compact-space replacement of the flat-space momentum integral: the density of states is not approximated, but given exactly by the degeneracy $n^2$.

Introducing dimensionless variables
\begin{equation}
\chi=k^{-1/2}\phi,
\qquad
u_k(\chi)=k^{-3}U_k(\phi),
\qquad
\tau=ka ,
\end{equation}
the flow becomes
\begin{equation}
\partial_tu_k(\chi)
=
-3u_k(\chi)
+
\frac{1}{2}\chi u_k'(\chi)
+
\frac{\tau^{2m-3}}{2\pi^2}
\sum_{n=1}^{\infty}
\frac{n^2}
{\left[n^2+M^2(\chi)\right]^m},
\label{eq:dimensionlessflow}
\end{equation}
with
\begin{equation}
M^2(\chi)
=
\tau^2[1+u_k''(\chi)]-1 .
\label{eq:Mdimensionless}
\end{equation}
At fixed $m$ and $\tau$, Eq.~\eqref{eq:dimensionlessflow} defines an autonomous compact spectral flow.  The geometry enters only through the spectral gaps, degeneracies, and the intrinsic location of the cutoff.

The spectral thresholds are meromorphic functions generated by
\begin{equation}
\mathcal{F}(\alpha)=
\sum_{n=1}^{\infty}\frac{1}{n^2+\alpha}
=
\frac{\pi\sqrt{\alpha}\coth(\pi\sqrt{\alpha})-1}{2\alpha}.
\label{eq:Falpha}
\end{equation}
For $m\geq2$,
\begin{equation}
\sum_{n=1}^{\infty}
\frac{n^2}{(n^2+\alpha)^m}
=
\frac{(-1)^{m-2}}{(m-2)!}
\left[
\partial_\beta^{m-2}
\left(
\mathcal{F}+\frac{\alpha}{m-1}\mathcal{F}'
\right)
\right]_{\beta=\alpha},
\label{eq:threshold_general}
\end{equation}
where $\mathcal{F}$ and $\mathcal{F}'$ are evaluated at $\beta$.  The poles occur at $n^2+\alpha=0$, with $\alpha=M^2(\chi)$ in the flow.

We work in the LPA, set the anomalous dimension to zero, and expand around the origin.  The crucial question is whether the compact spectral flow reproduces the flat-space fixed-point structure, and how strongly this structure depends on the finite spectral resolution $\tau$.

For all truncations and regulator parameters studied, the answer is affirmative.  The non-Gaussian fixed point persists on $\Sthree$, has one relevant direction, and approaches the flat-space proper-time LPA fixed point as $\tau$ increases.  At finite $\tau$ the critical exponents receive corrections associated with the discreteness of the low-lying spectrum, but these corrections remain small.  For the broadest threshold considered, $m=2$, the variation of $\nu$ is at the percent level.  For more localized thresholds it rapidly decreases, reaching the $10^{-3}$--$10^{-4}$ level in the $N=2$ truncation for $m=8,12$.  The leading irrelevant exponent $\omega$ is more sensitive to the truncation, but shows the same flattening with increasing $\tau$.

\begin{figure}[t]
\centering
\includegraphics[width=\columnwidth]{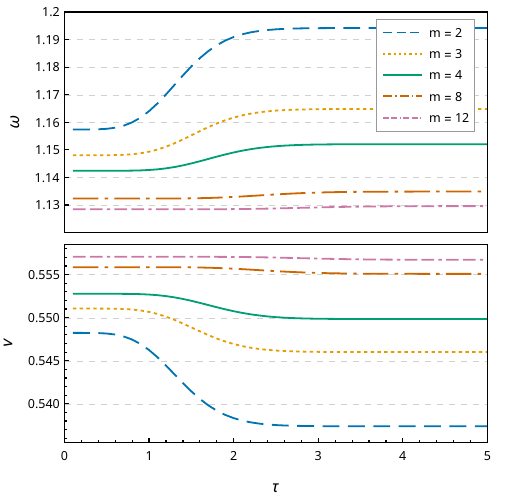}
\caption{
Intrinsic-resolution dependence of the critical exponents in the $N=2$ truncation.  Both $\nu$ and $\omega$ approach their flat-space values as $\tau$ increases.  The residual finite-resolution dependence decreases as the proper-time threshold is made more localized by increasing $m$.
}
\label{fig:exponents}
\end{figure}

Figure~\ref{fig:exponents} displays this behavior.  The flattening of the curves as $\tau$ grows is the compact-space analogue of the thermodynamic limit: the spectral density becomes effectively continuous and the homogeneous mode contributes a negligible fraction of the modes sampled by the flow.  Conversely, at $\tau=O(1)$ the cutoff acts directly on the first global modes, and the flow retains finite-size and curvature information through the exact gaps and degeneracies.

The dependence on the proper-time parameter has a simple interpretation.  A broad kernel (small $m$) samples a wider range of eigenvalues and is therefore more sensitive to the detailed structure of the low-lying spectrum.  Increasing $m$ localizes the threshold around the selected spectral shell, reducing the sensitivity to modes away from the cutoff. Thus the finite-$\tau$ variation has two aspects: its existence reflects the physical finite-resolution spectrum of the compact space, while its  magnitude is controlled by the regulator choice and by the truncation. The flattening of the curves at large $\tau$ indicates that the compact spectral flow approaches the regulator-specific flat-space LPA limit as the spectrum becomes dense.

The main result is therefore not merely that the Wilson--Fisher fixed point can be computed on $\Sthree$, but that it is recovered without introducing momentum shells.  The fixed point is already present in the compact spectral problem, while the flat-space fixed point appears as the dense-spectrum limit.  Compactness and curvature deform the approximate flow through gaps, degeneracies, and the homogeneous mode, but they do not change its fixed-point physical content.

\paragraph*{Conclusion.}
We conclude that Wilsonian coarse graining does not require momentum as a fundamental notion.  What is required is an invariant ordering of fluctuations by resolvability. On a compact curved space this ordering is supplied by the spectrum of the covariant kinetic operator.
In this sense the intrinsic cutoff is also relational \cite{ Tambornino:2011vg,Goeller:2022rsx,Baldazzi:2021fye,Ferrero:2024rvi}: its physical meaning is fixed from within, by comparing the RG resolution with the geometry on which the field propagates.
This parallels finite-temperature and finite-volume RG, where the cutoff scale becomes meaningful through dimensionless ratios such as $k\beta$ or $kL$ rather than through an external reference scale~\cite{Brezin:1981gm,Brezin:1985xx,Privman:1984zz,Cardy:1988}.
The present construction provides a minimal test bed for such momentum-free Wilsonian RG.   Extensions beyond LPA, to other compact manifolds and to fields with spin, will determine how broadly intrinsic spectral coarse graining can be used in  quantum field theory in curved spacetime, quantum gravity and asymptotic safety \cite{Reuter:1996cp, Percacci:2017fkn,Reuter:2019byg}.

\paragraph{Acknowledgments.}
We would like to thank Roberto Percacci for useful discussions.

\bibliographystyle{apsrev4-2}
\bibliography{literature}

@article{Bonanno:2019ukb,
    author = "Bonanno, Alfio and Lippoldt, Stefan and Percacci, Roberto and Vacca, Gian Paolo",
    title = "{On Exact Proper Time Wilsonian RG Flows}",
    eprint = "1912.08135",
    archivePrefix = "arXiv",
    primaryClass = "hep-th",
    doi = "10.1140/epjc/s10052-020-7798-9",
    journal = "Eur. Phys. J. C",
    volume = "80",
    number = "3",
    pages = "249",
    year = "2020"
}

@article{Litim:2010tt,
    author = "Litim, Daniel F. and Zappala, Dario",
    title = "{Ising exponents from the functional renormalisation group}",
    eprint = "1009.1948",
    archivePrefix = "arXiv",
    primaryClass = "hep-th",
    doi = "10.1103/PhysRevD.83.085009",
    journal = "Phys. Rev. D",
    volume = "83",
    pages = "085009",
    year = "2011"
}

@article{Kadanoff:1966wm,
  author = {Kadanoff, Leo P.},
  title = {Scaling laws for Ising models near $T_c$},
  journal = {Physics Physique Fizika},
  volume = {2},
  pages = {263--272},
  year = {1966},
  doi = {10.1103/PhysicsPhysiqueFizika.2.263}
}

@article{Wilson:1971dc,
  author = {Wilson, Kenneth G.},
  title = {Renormalization Group and Critical Phenomena. I. Renormalization Group and the Kadanoff Scaling Picture},
  journal = {Phys. Rev. B},
  volume = {4},
  pages = {3174--3183},
  year = {1971},
  doi = {10.1103/PhysRevB.4.3174}
}

@article{Wilson:1971bg,
  author = {Wilson, Kenneth G.},
  title = {Renormalization Group and Critical Phenomena. II. Phase-Space Cell Analysis of Critical Behavior},
  journal = {Phys. Rev. B},
  volume = {4},
  pages = {3184--3205},
  year = {1971},
  doi = {10.1103/PhysRevB.4.3184}
}

@article{Wilson:1972zz,
  author = {Wilson, Kenneth G. and Fisher, Michael E.},
  title = {Critical Exponents in 3.99 Dimensions},
  journal = {Phys. Rev. Lett.},
  volume = {28},
  pages = {240--243},
  year = {1972},
  doi = {10.1103/PhysRevLett.28.240}
}

@article{Wilson:1974mb,
  author = {Wilson, Kenneth G. and Kogut, John B.},
  title = {The Renormalization Group and the epsilon Expansion},
  journal = {Phys. Rept.},
  volume = {12},
  pages = {75--199},
  year = {1974},
  doi = {10.1016/0370-1573(74)90023-4}
}

@book{ZinnJustin:2002ru,
  author = {Zinn-Justin, Jean},
  title = {Quantum Field Theory and Critical Phenomena},
  publisher = {Oxford University Press},
  edition = {4},
  year = {2002}
}

@article{Pelissetto:2000ek,
  author = {Pelissetto, Andrea and Vicari, Ettore},
  title = {Critical phenomena and renormalization-group theory},
  journal = {Phys. Rept.},
  volume = {368},
  pages = {549--727},
  year = {2002},
  eprint = {cond-mat/0012164},
  archivePrefix = {arXiv}
}

@article{Wegner:1973,
  author = {Wegner, F. J. and Houghton, A.},
  title = {Renormalization Group Equation for Critical Phenomena},
  journal = {Phys. Rev. A},
  volume = {8},
  pages = {401--412},
  year = {1973},
  doi = {10.1103/PhysRevA.8.401}
}

@article{Polchinski:1983gv,
  author = {Polchinski, Joseph},
  title = {Renormalization and Effective Lagrangians},
  journal = {Nucl. Phys. B},
  volume = {231},
  pages = {269--295},
  year = {1984},
  doi = {10.1016/0550-3213(84)90287-6}
}

@article{Hasenfratz:1986dm,
  author = {Hasenfratz, Anna and Hasenfratz, Peter},
  title = {Renormalization Group Study of Scalar Field Theories},
  journal = {Nucl. Phys. B},
  volume = {270},
  pages = {687--701},
  year = {1986}
}

@article{Wetterich:1992yh,
  author = {Wetterich, Christof},
  title = {Exact evolution equation for the effective potential},
  journal = {Phys. Lett. B},
  volume = {301},
  pages = {90--94},
  year = {1993},
  doi = {10.1016/0370-2693(93)90726-X}
}

@article{Morris:1993qb,
  author = {Morris, Tim R.},
  title = {The exact renormalization group and approximate solutions},
  journal = {Int. J. Mod. Phys. A},
  volume = {9},
  pages = {2411--2450},
  year = {1994},
  doi = {10.1142/S0217751X94000972},
  eprint = {hep-ph/9308265},
  archivePrefix = {arXiv}
}

@article{Berges:2000ew,
  author = {Berges, Juergen and Tetradis, Nikolaos and Wetterich, Christof},
  title = {Non-perturbative renormalization flow in quantum field theory and statistical physics},
  journal = {Phys. Rept.},
  volume = {363},
  pages = {223--386},
  year = {2002},
  doi = {10.1016/S0370-1573(01)00098-9},
  eprint = {hep-ph/0005122},
  archivePrefix = {arXiv}
}

@article{Delamotte:2012nb,
  author = {Delamotte, Bertrand},
  title = {An Introduction to the Nonperturbative Renormalization Group},
  journal = {Lect. Notes Phys.},
  volume = {852},
  pages = {49--132},
  year = {2012},
  doi = {10.1007/978-3-642-27320-9_2},
  eprint = {cond-mat/0702365},
  archivePrefix = {arXiv}
}

@article{Dupuis:2020fhh,
  author = {Dupuis, Nicolas and Canet, Leonie and Eichhorn, Astrid and Metzner, Walter and Pawlowski, Jan M. and Tissier, Matthieu and Wschebor, Nicolas},
  title = {The nonperturbative functional renormalization group and its applications},
  journal = {Phys. Rept.},
  volume = {910},
  pages = {1--114},
  year = {2021},
  doi = {10.1016/j.physrep.2021.01.001},
  eprint = {2006.04853},
  archivePrefix = {arXiv},
  primaryClass = {cond-mat.stat-mech}
}

@article{Tetradis:1993ts,
  author = {Tetradis, Nikolaos and Wetterich, Christof},
  title = {Critical exponents from the effective average action},
  journal = {Nucl. Phys. B},
  volume = {422},
  pages = {541--592},
  year = {1994},
  eprint = {hep-ph/9308214},
  archivePrefix = {arXiv}
}

@article{Morris:1994ie,
  author = {Morris, Tim R.},
  title = {Derivative expansion of the exact renormalization group},
  journal = {Phys. Lett. B},
  volume = {329},
  pages = {241--248},
  year = {1994},
  eprint = {hep-ph/9403340},
  archivePrefix = {arXiv}
}

@article{Litim:2001up,
  author = {Litim, Daniel F.},
  title = {Optimized renormalization group flows},
  journal = {Phys. Rev. D},
  volume = {64},
  pages = {105007},
  year = {2001},
  doi = {10.1103/PhysRevD.64.105007},
  eprint = {hep-th/0103195},
  archivePrefix = {arXiv}
}

@article{Litim:2002cf,
  author = {Litim, Daniel F.},
  title = {Critical exponents from optimized renormalization group flows},
  journal = {Nucl. Phys. B},
  volume = {631},
  pages = {128--158},
  year = {2002},
  doi = {10.1016/S0550-3213(02)00186-4},
  eprint = {hep-th/0203006},
  archivePrefix = {arXiv}
}

@article{Hollands:2002ux,
    author = "Hollands, Stefan and Wald, Robert M.",
    title = "{On the renormalization group in curved space-time}",
    eprint = "gr-qc/0209029",
    archivePrefix = "arXiv",
    doi = "10.1007/s00220-003-0837-1",
    journal = "Commun. Math. Phys.",
    volume = "237",
    pages = "123--160",
    year = "2003"
}

@book{Thiemann:2007pyv,
    author = "Thiemann, Thomas",
    title = "{Modern Canonical Quantum General Relativity}",
    doi = "10.1017/CBO9780511755682",
    isbn = "978-0-511-75568-2, 978-0-521-84263-1",
    publisher = "Cambridge University Press",
    series = "Cambridge Monographs on Mathematical Physics",
    year = "2007"
}

@article{Canet:2003qd,
  author = {Canet, Leonie and Delamotte, Bertrand and Mouhanna, Dominique and Vidal, Julien},
  title = {Optimization of the derivative expansion in the nonperturbative renormalization group},
  journal = {Phys. Rev. D},
  volume = {67},
  pages = {065004},
  year = {2003},
  eprint = {hep-th/0211055},
  archivePrefix = {arXiv}
}

@article{Fisher:1972zza,
  author = {Fisher, Michael E. and Barber, Michael N.},
  title = {Scaling Theory for Finite-Size Effects in the Critical Region},
  journal = {Phys. Rev. Lett.},
  volume = {28},
  pages = {1516--1519},
  year = {1972},
  doi = {10.1103/PhysRevLett.28.1516}
}

@article{Tambornino:2011vg,
  author = {Tambornino, Johannes},
  title = {Relational Observables in Gravity: a Review},
  journal = {SIGMA},
  volume = {8},
  pages = {017},
  year = {2012},
  eprint = {1109.0740},
  archivePrefix = {arXiv},
  doi = {10.3842/SIGMA.2012.017}
}

@book{Cardy:1988,
  editor = {Cardy, John L.},
  title = {Finite-Size Scaling},
  publisher = {North-Holland},
  address = {Amsterdam},
  year = {1988}
}

@article{Brezin:1981gm,
  author = {Brezin, E.},
  title = {An Investigation of Finite Size Scaling},
  journal = {J. Phys. France},
  volume = {43},
  pages = {15--22},
  year = {1982},
  doi = {10.1051/jphys:0198200430101500}
}

@article{Brezin:1985xx,
  author = {Brezin, E. and Zinn-Justin, J.},
  title = {Finite size effects in phase transitions},
  journal = {Nucl. Phys. B},
  volume = {257},
  pages = {867--893},
  year = {1985}
}

@article{Ferrero:2024rvi,
    author = "Ferrero, Renata and Thiemann, Thomas",
    title = "{Relational Lorentzian Asymptotically Safe Quantum Gravity: Showcase Model}",
    eprint = "2404.18224",
    archivePrefix = "arXiv",
    primaryClass = "hep-th",
    doi = "10.3390/universe10110410",
    journal = "Universe",
    volume = "10",
    number = "11",
    pages = "410",
    year = "2024"
}

@article{Baldazzi:2021fye,
    author = "Baldazzi, Alessio and Falls, Kevin and Ferrero, Renata",
    title = "{Relational observables in asymptotically safe gravity}",
    eprint = "2112.02118",
    archivePrefix = "arXiv",
    primaryClass = "hep-th",
    doi = "10.1016/j.aop.2022.168822",
    journal = "Annals Phys.",
    volume = "440",
    pages = "168822",
    year = "2022"
}

@article{Benedetti:2014gja,
    author = "Benedetti, Dario",
    title = "{Critical behavior in spherical and hyperbolic spaces}",
    eprint = "1403.6712",
    archivePrefix = "arXiv",
    primaryClass = "cond-mat.stat-mech",
    reportNumber = "AEI-2014-008",
    doi = "10.1088/1742-5468/2015/01/P01002",
    journal = "J. Stat. Mech.",
    volume = "1501",
    pages = "P01002",
    year = "2015"
}

@article{Ferrero:2025ugd,
    author = "Ferrero, Renata and Naso, Vincenzo and Percacci, Roberto",
    title = "{Quantum Fields and the Cosmological Constant}",
    eprint = "2503.17203",
    archivePrefix = "arXiv",
    primaryClass = "hep-th",
    doi = "10.3390/universe11060173",
    journal = "Universe",
    volume = "11",
    number = "6",
    pages = "173",
    year = "2025"
}

@article{Banerjee:2023ztr,
    author = "Banerjee, Rudrajit and Becker, Maximilian and Ferrero, Renata",
    title = "{N-cutoff regularization for fields on hyperbolic space}",
    eprint = "2302.03547",
    archivePrefix = "arXiv",
    primaryClass = "hep-th",
    reportNumber = "MITP-23-003",
    doi = "10.1103/PhysRevD.109.025008",
    journal = "Phys. Rev. D",
    volume = "109",
    number = "2",
    pages = "025008",
    year = "2024"
}

@article{Becker:2021pwo,
    author = "Becker, Maximilian and Reuter, Martin",
    title = "{Background independent field quantization with sequences of gravity-coupled approximants. II. Metric fluctuations}",
    eprint = "2109.09496",
    archivePrefix = "arXiv",
    primaryClass = "hep-th",
    doi = "10.1103/PhysRevD.104.125008",
    journal = "Phys. Rev. D",
    volume = "104",
    number = "12",
    pages = "125008",
    year = "2021"
}

@article{Chen_2000,
  author = {Chen, X. S. and Dohm, V.},
  title = {Relation between bulk order-parameter correlation function and finite-size scaling},
  journal = {Eur. Phys. J. B},
  volume = {15},
  pages = {283--296},
  year = {2000},
  doi = {10.1007/s100510051127}
}

@article{Fister:2015eca,
  author = {Fister, Leonard and Pawlowski, Jan Martin},
  title = {Functional renormalization group in a finite volume},
  journal = {Phys. Rev. D},
  volume = {92},
  pages = {076009},
  year = {2015},
  doi = {10.1103/PhysRevD.92.076009},
  eprint = {1504.05166},
  archivePrefix = {arXiv},
  primaryClass = {hep-ph}
}

@article{Liao:1996pd,
  author = {Liao, S.-B.},
  title = {On connection between momentum cutoff and the proper time regularizations},
  journal = {Phys. Rev. D},
  volume = {53},
  pages = {2020--2036},
  year = {1996},
  doi = {10.1103/PhysRevD.53.2020},
  eprint = {hep-th/9501124},
  archivePrefix = {arXiv}
}

@article{Bonanno:2000yp,
  author = {Bonanno, Alfio and Zappala, Dario},
  title = {Towards an accurate determination of the critical exponents with the renormalization group flow equations},
  journal = {Phys. Lett. B},
  volume = {504},
  pages = {181--187},
  year = {2001},
  doi = {10.1016/S0370-2693(01)00286-1},
  eprint = {hep-th/0010095},
  archivePrefix = {arXiv}
}

@article{Bonanno:2004sy,
  author = {Bonanno, Alfio and Zappala, Dario},
  title = {Proper time flow equation and critical exponents},
  journal = {Nucl. Phys. B},
  volume = {693},
  pages = {501--511},
  year = {2004},
  eprint = {hep-th/0402039},
  archivePrefix = {arXiv}
}

@article{Schwinger:1951nm,
  author = {Schwinger, Julian S.},
  title = {On gauge invariance and vacuum polarization},
  journal = {Phys. Rev.},
  volume = {82},
  pages = {664--679},
  year = {1951}
}

@book{DeWitt:1965,
  author = {DeWitt, Bryce S.},
  title = {Dynamical Theory of Groups and Fields},
  publisher = {Gordon and Breach},
  address = {New York},
  year = {1965}
}

@book{BirrellDavies:1982,
  author = {Birrell, N. D. and Davies, P. C. W.},
  title = {Quantum Fields in Curved Space},
  publisher = {Cambridge University Press},
  year = {1982}
}

@book{Reuter:2019byg,
    author = "Reuter, Martin and Saueressig, Frank",
    title = "{Quantum Gravity and the Functional Renormalization Group}: {The Road towards Asymptotic Safety}",
    isbn = "978-1-107-10732-8, 978-1-108-67074-6",
    publisher = "Cambridge University Press",
    month = "1",
    year = "2019"
}

@book{ParkerToms:2009,
  author = {Parker, Leonard and Toms, David},
  title = {Quantum Field Theory in Curved Spacetime: Quantized Fields and Gravity},
  publisher = {Cambridge University Press},
  year = {2009}
}

@book{Buchbinder:1992rb,
  author = {Buchbinder, I. L. and Odintsov, S. D. and Shapiro, I. L.},
  title = {Effective Action in Quantum Gravity},
  publisher = {IOP Publishing},
  address = {Bristol},
  year = {1992}
}

@article{Vassilevich:2003xt,
  author = {Vassilevich, Dmitri V.},
  title = {Heat kernel expansion: User's manual},
  journal = {Phys. Rept.},
  volume = {388},
  pages = {279--360},
  year = {2003},
  doi = {10.1016/j.physrep.2003.09.002},
  eprint = {hep-th/0306138},
  archivePrefix = {arXiv}
}

@book{Gilkey:1995,
  author = {Gilkey, Peter B.},
  title = {Invariance Theory, the Heat Equation, and the Atiyah-Singer Index Theorem},
  publisher = {CRC Press},
  edition = {2},
  year = {1995}
}

@book{Kirsten:2001,
  author = {Kirsten, Klaus},
  title = {Spectral Functions in Mathematics and Physics},
  publisher = {Chapman and Hall/CRC},
  year = {2001}
}

@book{Chavel:1984,
  author = {Chavel, Isaac},
  title = {Eigenvalues in Riemannian Geometry},
  publisher = {Academic Press},
  address = {Orlando},
  year = {1984}
}

@article{Camporesi:1994ga,
  author = {Camporesi, Roberto and Higuchi, Atsushi},
  title = {On the eigenfunctions of the Dirac operator on spheres and real hyperbolic spaces},
  journal = {J. Geom. Phys.},
  volume = {20},
  pages = {1--18},
  year = {1996},
  doi = {10.1016/0393-0440(95)00042-9},
  eprint = {gr-qc/9505009},
  archivePrefix = {arXiv}
}

@article{Reuter:1996cp,
  author = {Reuter, Martin},
  title = {Effective average actions and nonperturbative evolution equations},
  journal = {hep-th/9602012},
  year = {1996},
  eprint = {hep-th/9602012},
  archivePrefix = {arXiv}
}

@article{Goeller:2022rsx,
    author = "Goeller, Christophe and Hoehn, Philipp A. and Kirklin, Josh",
    title = "{Diffeomorphism-invariant observables and dynamical frames in gravity: reconciling bulk locality with general covariance}",
    journal = "",
    eprint = "2206.01193",
    archivePrefix = "arXiv",
    primaryClass = "hep-th",
    month = "6",
    year = "2022"
}

@book{Percacci:2017fkn,
    author = "Percacci, Robert",
    title = "{An Introduction to Covariant Quantum Gravity and Asymptotic Safety}",
    doi = "10.1142/10369",
    isbn = "978-981-320-717-2, 978-981-320-719-6",
    publisher = "World Scientific",
    series = "100 Years of General Relativity",
    volume = "3",
    year = "2017"
}

@article{Camporesi1990,
  author  = {Camporesi, Roberto},
  title   = {Harmonic Analysis and Propagators on Homogeneous Spaces},
  journal = {Physics Reports},
  volume  = {196},
  pages   = {1--134},
  year    = {1990},
  doi     = {10.1016/0370-1573(90)90120-Q}
}

@book{Kiefer:2007ria,
    author = "Kiefer, Claus",
    title = "{Quantum Gravity}",
    edition = "2nd ed",
    isbn = "978-0-19-921252-1",
    publisher = "Oxford University Press",
    address = "New York",
    year = "2007"
}

@article{Privman:1984zz,
    author = "Privman, Valdimir and Fisher, Michael E.",
    title = "{Universal critical amplitudes in finite-size scaling}",
    doi = "10.1103/PhysRevB.30.322",
    journal = "Phys. Rev. B",
    volume = "30",
    pages = "322--327",
    year = "1984"
}

@article{Becker:2020mjl,
    author = "Becker, Maximilian and Reuter, Martin",
    title = "{Background Independent Field Quantization with Sequences of Gravity-Coupled Approximants}",
    eprint = "2008.09430",
    archivePrefix = "arXiv",
    primaryClass = "gr-qc",
    doi = "10.1103/PhysRevD.102.125001",
    journal = "Phys. Rev. D",
    volume = "102",
    number = "12",
    pages = "125001",
    year = "2020"
}

\end{document}